\documentclass[9pt,a4paper,onecolumn,draftcls]{IEEEtran}
\usepackage{etex}

\usepackage[a4,frame,center]{crop}
\usepackage{url}
\usepackage{psfrag}
\usepackage[usenames,dvipsnames]{xcolor}
\usepackage{tikz}
\usepackage{graphicx}
\DeclareGraphicsExtensions{.eps}
\usepackage{balance}
\usepackage{tikz}
\usepackage{pstricks,pst-node}
\usepackage{amsmath}
\usepackage{subfig}
\usepackage{amssymb}
\usepackage{amsfonts}
\usepackage{psfrag}
\usepackage{amsmath,epsfig,amsfonts,amssymb,psfrag}
\usepackage{cite}
\usepackage{algorithm}
\usepackage{algorithmic}
\usepackage{pstricks,pst-node}
\usepackage{lscape}

\usetikzlibrary{patterns}
\usetikzlibrary{shapes}
\usetikzlibrary{shadows}
\usetikzlibrary{positioning}
\usetikzlibrary{arrows}
\usetikzlibrary{calc}

\usepackage{pstricks, pst-node, pst-plot, pst-circ}
\usepackage{moredefs}
\providelength{\AxesLineWidth}       
\providelength{\plotwidth}           
\providelength{\LineWidth}           
\providelength{\MarkerSize}          
\newrgbcolor{GridColor}{0.8 0.8 0.8}%

\ifCLASSINFOpdf
\else
\fi
\begin{document}
\title{
Interference Alignment: Practical Challenges and Test-bed Implementation
}

\author{
$^{\dagger}$Nima Najari Moghadam, $^{\dagger}$Hamed Farhadi,  $^{\dagger}$Per Zetterberg, $^{\star}$Majid Nasiri Khormuji, and $^{\dagger}$Mikael Skoglund\\
\IEEEauthorblockA{$^{\dagger}$
ACCESS Linnaeus Center,
KTH Royal Institute of Technology,
Stockholm, Sweden\\
Emails: \{nimanm, farhadih, perz, skoglund\}@ee.kth.se, \\
$^{\star}$Huawei Technologies Sweden AB, Stockholm, Sweden,
Email: majid.nk@huawei.com.
}
}
\maketitle

\section{Introduction}
\label{sec: introduction}
Data traffic over wireless communication networks has experienced a tremendous growth in the last decade,
and it is predicted to exponentially increase in the next decades \cite{Osseiran2013}.
Enabling future wireless networks to fulfill this expectation is a challenging task both due to the scarcity of radio resources (e.g. spectrum and energy), and also the inherent characteristics of the wireless transmission medium.
Wireless transmission is in general subject to two phenomena: \textit{fading} and \textit{interference}.
The former is a consequence of reflectors scattered in the environment surrounding a transmitter and a receiver such that the receiver observes a superposition of multiple copies of the transmitted signal.
The superposition of the signals can be either constructive or destructive depending on the phase shift and attenuation of the received signals from different paths.
The randomness of fading may degrade communication quality.
Several effective techniques have been developed to overcome the adverse effects of random fading.
For instance, multiple antennas transmission techniques are proposed to realize spatial diversity and to improve the performance of wireless systems in the presence of fading \cite{Tarokh1998}.
In addition, in multi-user networks because of the broadcast nature of wireless transmission medium,
each user's communication is interfered by other users.
Inter-user interference can severely degrade the communication quality and makes communication of different users interrelated; thus, finding the optimum interference management strategy becomes a challenging problem.

Conventional interference management strategies
including time-division multiple access (TDMA) and frequency-division multiple-access (FDMA) avoid the inter-user interference by allocating orthogonal resources in time and frequency to different users, respectively.
Interference is consequently avoided at the cost of low spectral efficiency.
Thus, it was believed that the performance of wireless networks is limited by interference in general.
However, the elegant \emph{interference alignment} concept
\cite{Maddah-Ali2008,Cadambe2008} reveals that with proper
transmission signalling design,
different interference signals can in fact be aligned together,
such that more radio resources can be assigned to the desired transmission.
For instance, in the case of a multi-user interference network with more than two source--destination pairs,
the interference signals at each destination can be aligned such that maximally half of the signal
space can be left to its desired signal \cite{Cadambe2008}.
Therefore, each user may achieve half of the interference-free transmission rate no matter how many interferers exist \cite{Cadambe2008}.
Although interference alignment can achieve a larger data rate compared to orthogonal transmission strategies,
several challenges should be addressed to enable the deployment of this technique in future wireless networks \cite{Ayach2013,Osseiran2013}.
For instance, to perform interference alignment, normally, global channel state information (CSI) is required to be perfectly known at all terminals.
Clearly, acquiring such channel knowledge is a challenging problem in practice and proper channel training and channel state feedback techniques need to be deployed.
In addition, since the channels are time-varying proper adaptive transmission is needed.

To investigate whether the outstanding performance of signal processing algorithms inspired
by interference alignment can be preserved in real environment, practical verifications is needed.
Wireless test-beds (e.g. the ones based on USRP or WARP)
can be used as a platform for the experimental verification of the novel interference management algorithms.

This chapter review recent advances in practical aspects of interference alignment.
It also presents recent test-bed implementations of signal processing algorithms for the realization of interference alignment.
In Section \ref{Review sec: Interefernce Alignment} we give a brief overview on the interference alignment concept.
Section \ref{Sec:Practical Challenges of Interference Alignment} presents the structure of a canonical transmitter and receiver to realize interference alignment, and discuss channel training and channel state feedback for these systems.
A brief review on test-bed implementations of interference alignment solutions is presented in Section \ref{Sec: Review on Test-bed Implementation of Interference Alignment}. Section \ref{Sec: KTH four-multi Test-bed Setup}, introduces hardware and software setup of the test-bed used in this chapter for implementation of interference alignment.
The test-bed implementation of iterative transceiver design and power control algorithm is presented in Section \ref{Sec: Test-bed Implementation of the Iterative Transceiver Design and Power Control}.
We discuss the test-bed implementation of compressed feedback scheme for interference alignment scheme in Section \ref{Sec: Test-bed Implementation of the Interference Alignment with Compressed Feedback}.
Finally, Section \ref{Sec: Conclusion} concludes the chapter.

\section{$K$-user ($K>2$) Interference Networks}\label{Review sec: Interefernce Alignment}
\begin{figure}
    \centering
    \scalebox{0.4}
    {


\begin{tikzpicture}[>=latex']
\tikzset{Source/.style={rectangle, draw, very thick, minimum width=1.6cm, minimum height=1cm, rounded corners=2mm}}
\tikzset{Destination/.style={rectangle, draw, very thick, minimum width=1.6cm, minimum height=1cm, rounded corners=2mm}}
\tikzset{Noise/.style={circle, draw, thick, minimum size=4mm}}
\tikzset{descr/.style={fill=white}}
\Large

\node[Source] (S3) at (-0.5,0) {$\textnormal{S}_K$};
\node[Source] (S2) at (-0.5,1.25) {$\textnormal{S}_2$};
\node[Source] (S1) at (-0.5,2.5) {$\textnormal{S}_1$};


\node[Destination] (D3) at (5,0) {$\textnormal{D}_K$};
\node[Destination] (D2) at (5,1.25) {$\textnormal{D}_2$};
\node[Destination] (D1) at (5,2.5) {$\textnormal{D}_1$};


\draw [rounded corners=2mm,very thick] (1.25,-0.5) rectangle (3.25,3);
\draw node at (2.25,1.25) {Channel};
\draw node at (2.25,-1.25) {(a)};

\draw [->] (S1.east) -- (1.25,2.5);
\draw [->] (S2.east) -- (1.25,1.25);
\draw [->] (S3.east) -- (1.25,0);

\draw [->] (3.25,2.5) -- (D1.west);
\draw [->] (3.25,1.25) -- (D2.west);
\draw [->] (3.25,0) -- (D3.west);

\draw [red,pattern color=red, pattern=horizontal lines] ($(S3)+(-6.5,-0.3)$) rectangle ($(S3)+(-2,0.3)$) ;

\draw [green,pattern color=green, pattern=north east lines] ($(S2)+(-6.5,-0.3)$) rectangle ($(S2)+(-2,0.3)$) ;

\draw [blue,pattern color=blue, pattern=vertical lines] ($(S1)+(-6.5,-0.3)$) rectangle ($(S1)+(-2,0.3)$);

\draw [red,pattern color=red, pattern=horizontal lines]
($(D3)+(2,-0.3)$) rectangle ($(D3)+(6.5,0.3)$);
\draw [green,pattern color=green, pattern=north east lines] ($(D3)+(2,-0.3)$) rectangle ($(D3)+(6.5,0.3)$);
\draw [blue,pattern color=blue, pattern=vertical lines]($(D3)+(2,-0.3)$) rectangle ($(D3)+(6.5,0.3)$);

\draw [red,pattern color=red, pattern=horizontal lines] ($(D2)+(2,-0.3)$) rectangle ($(D2)+(6.5,0.3)$);
\draw [green,pattern color=green, pattern=north east lines] ($(D2)+(2,-0.3)$) rectangle ($(D2)+(6.5,0.3)$);
\draw [blue,pattern color=blue, pattern=vertical lines] ($(D2)+(2,-0.3)$) rectangle ($(D2)+(6.5,0.3)$);

\draw [red,pattern color=red, pattern=horizontal lines] ($(D1)+(2,-0.3)$) rectangle ($(D1)+(6.5,0.3)$);
\draw [green,pattern color=green, pattern=north east lines] ($(D1)+(2,-0.3)$) rectangle ($(D1)+(6.5,0.3)$);
\draw [blue,pattern color=blue, pattern=vertical lines] ($(D1)+(2,-0.3)$) rectangle ($(D1)+(6.5,0.3)$);


\node[Source] (S3_TDMA) at (-0.5,-5.5) {$\textnormal{S}_K$};
\node[Source] (S2_TDMA) at (-0.5,-4.25) {$\textnormal{S}_2$};
\node[Source] (S1_TDMA) at (-0.5,-3) {$\textnormal{S}_1$};


\node[Destination] (D3_TDMA) at (5,-5.5) {$\textnormal{D}_K$};
\node[Destination] (D2_TDMA) at (5,-4.25) {$\textnormal{D}_2$};
\node[Destination] (D1_TDMA) at (5,-3) {$\textnormal{D}_1$};


\draw [rounded corners=2mm,very thick] (1.25,-6) rectangle (3.25,-2.5);
\draw node at (2.25,-4.25) {Channel};
\draw node at (2.25,-6.75) {(b)};

\draw [->] (S1_TDMA.east) -- (1.25,-3);
\draw [->] (S2_TDMA.east) -- (1.25,-4.25);
\draw [->] (S3_TDMA.east) -- (1.25,-5.5);

\draw [->] (3.25,-3) -- (D1_TDMA.west);
\draw [->] (3.25,-4.25) -- (D2_TDMA.west);
\draw [->] (3.25,-5.5) -- (D3_TDMA.west);

\draw [red,pattern color=red, pattern=horizontal lines] ($(S3_TDMA)+(-3.5,-0.3)$) rectangle ($(S3_TDMA)+(-2,0.3)$) ;

\draw [green,pattern color=green, pattern=north east lines] ($(S2_TDMA)+(-5,-0.3)$) rectangle ($(S2_TDMA)+(-3.5,0.3)$) ;

\draw [blue,pattern color=blue, pattern=vertical lines] ($(S1_TDMA)+(-5,-0.3)$) rectangle ($(S1_TDMA)+(-6.5,0.3)$) ;

\draw [red,pattern color=red, pattern=horizontal lines] ($(D3_TDMA)+(5,-0.3)$) rectangle ($(D3_TDMA)+(6.5,0.3)$) ;

\draw [green,pattern color=green, pattern=north east lines] ($(D2_TDMA)+(3.5,-0.3)$) rectangle ($(D2_TDMA)+(5,0.3)$) ;

\draw [blue,pattern color=blue, pattern=vertical lines] ($(D1_TDMA)+(2,-0.3)$) rectangle ($(D1_TDMA)+(3.5,0.3)$) ;


\node[Source] (S3_IA) at (-0.5,-11) {$\textnormal{S}_K$};
\node[Source] (S2_IA) at (-0.5,-9.75) {$\textnormal{S}_2$};
\node[Source] (S1_IA) at (-0.5,-8.5) {$\textnormal{S}_1$};


\node[Destination] (D3_IA) at (5,-11) {$\textnormal{D}_K$};
\node[Destination] (D2_IA) at (5,-9.75) {$\textnormal{D}_2$};
\node[Destination] (D1_IA) at (5,-8.5) {$\textnormal{D}_1$};


\draw [rounded corners=2mm,very thick] (1.25,-11.5) rectangle (3.25,-8);
\draw node at (2.25,-9.75) {Channel};
\draw node at (2.25,-12.25) {(c)};

\draw [->] (S1_IA.east) -- (1.25,-8.5);
\draw [->] (S2_IA.east) -- (1.25,-9.75);
\draw [->] (S3_IA.east) -- (1.25,-11);

\draw [->] (3.25,-8.5) -- (D1_IA.west);
\draw [->] (3.25,-9.75) -- (D2_IA.west);
\draw [->] (3.25,-11) -- (D3_IA.west);

\draw [red,pattern color=red, pattern=horizontal lines] ($(S3_IA)+(-6.5,-0.3)$) rectangle ($(S3_IA)+(-4.25,0.3)$) ;

\draw [green,pattern color=green, pattern=north east lines] ($(S2_IA)+(-6.5,-0.3)$) rectangle ($(S2_IA)+(-4.25,0.3)$) ;

\draw [blue,pattern color=blue, pattern=vertical lines] ($(S1_IA)+(-6.5,-0.3)$) rectangle ($(S1_IA)+(-4.25,0.3)$) ;

\draw [red,pattern color=red, pattern=horizontal lines] ($(D3_IA)+(2,-0.3)$) rectangle ($(D3_IA)+(4.25,0.3)$) ;
\draw [green,pattern color=green, pattern=north east lines] ($(D3_IA)+(4.25,-0.3)$) rectangle ($(D3_IA)+(6,0.3)$) ;
\draw [blue,pattern color=blue, pattern=vertical lines] ($(D3_IA)+(4.25,-0.3)$) rectangle ($(D3_IA)+(6,0.3)$) ;

\draw [green,pattern color=green, pattern=north east lines]  ($(D2_IA)+(2,-0.3)$) rectangle ($(D2_IA)+(4.25,0.3)$) ;
\draw [red,pattern color=red, pattern=horizontal lines] ($(D2_IA)+(4.25,-0.3)$) rectangle ($(D2_IA)+(6,0.3)$) ;
\draw [blue,pattern color=blue, pattern=vertical lines] ($(D2_IA)+(4.25,-0.3)$) rectangle ($(D2_IA)+(6,0.3)$) ;

\draw [blue,pattern color=blue, pattern=horizontal lines]  ($(D1_IA)+(2,-0.3)$) rectangle ($(D1_IA)+(4.25,0.3)$) ;
\draw [green,pattern color=green, pattern=north east lines] ($(D1_IA)+(4.25,-0.3)$) rectangle ($(D1_IA)+(6,0.3)$) ;
\draw [red,pattern color=red, pattern=vertical lines] ($(D1_IA)+(4.25,-0.3)$) rectangle ($(D1_IA)+(6,0.3)$) ;

\end{tikzpicture}

    }
    \caption[Interference management schemes]{\footnotesize Transmission schemes in three-user interference networks: (a) non-orthogonal transmission and decdoing by treating interference as noise, (b) orthogonal transmission, and (c) interference alignment.}
    \label{Fig: Review_concept_all}
 \end{figure}
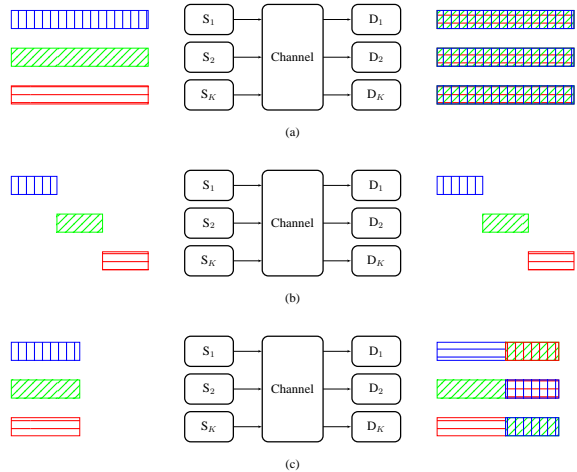
Consider a $K$-user $(K>2)$ interference network.
Several techniques have been devised for interference management in multi-user interference networks.
Three major approaches to deal with interference are illustrated in Fig.~\ref{Fig: Review_concept_all}.
In Fig.~\ref{Fig: Review_concept_all} (a)
all sources simultaneously transmit their signals in the same frequency band.
Each source applies the conventional single-user coding technique.
At each destination, the desired signal is superimposed by interference signals.
The destination performs decoding by treating the interference signals as noise.
This strategy is reasonable for cases that the receiver only knows the transmitted codebooks of the intended source.
In the low signal-to-noise ratio (SNR) region,
the level of interference can be controlled by proper power control techniques.
However, in the high-SNR regime, inter-user interference
is dominant.
Therefore, the power control alone is not sufficient to
manage the interference and this transmission strategy may not lead to a desirable performance.

The conventional approach to avoid interference at destinations is to orthogonalize the transmissions of different users.
Each source--destination pair has access to only a portion of the available channel, as shown in Fig.~\ref{Fig: Review_concept_all} (b).
Although signal reception at each destination does not
directly suffer from inter-user interference,
this scheme is not spectrally efficient due to the fact the resource; i.e. time or bandwidth, are divided among the source--destination pairs.
From Fig.~\ref{Fig: Review_concept_all} (b), we see that the interference signals span a large dimension of the received signal space at each destination to ensure orthogonal reception. However, if at each destination the dimension of the subspace occupied by only the interference signals can be reduced,
a larger interference-free subspace would be left for desired transmission.
This can be realized using a new technique called \textit{interference alignment} \cite{Maddah-Ali2008}. Specifically, interference alignment for interference networks refers to ``\textit{a construction of signals in such a manner
that they cast overlapping shadows at the receivers where they
constitute interference while they remain distinguishable at the
intended receivers where they are desired}'' \cite{Cadambe2008}.
In general, two conditions should be satisfied to perform interference alignment technique.
First, interference signals should be aligned at the same subspace, termed \textit{interference subspace}.
Next, we need to check whether the subspace left for the desired signal, called \textit{desired subspace},
is independent from the interference subspace.
The conditions are essentially required for the realization of the interference alignment techniques.
An illustrative representation of this concept is shown in Fig.~\ref{Fig: Review_concept_all} (c).

Interference alignment can be realized in different domains such
as space (across multiple antennas \cite{Maddah-Ali2008},
\cite{Cadambe2008}), time (exploiting propagation delays
\cite{Maleki2012,Maddah-Ali2010} or coding across
time-varying channels \cite{Cadambe2008}, \cite{Nazer2009}),
frequency (coding across different carriers in frequency-selective
channels \cite{BRA:12})
, and code (aligning interference in signal levels \cite{Motahari2009a}).
Combinations of domains can also be used e.g. space and frequency,
\cite{BRA:13}.
In the following, we briefly introduce \emph{Degrees of Freedom (DoF)} which is a performance measure for wireless networks at high-SNR regime.

\subsection{Degrees of Freedom Region}
Consider the $K$-user interference network in Fig.~\ref{Fig: Review_concept_all}. Source $\mathrm{S}_k$ ($k\in\{1,2,...,K\}$) intends to
send an independent message $w_k\in \mathcal{W}_k$
to its destination, where $\mathcal{W}_k$
denotes the corresponding message set.
The message $w_k$ is encoded to a
codeword of length $n$.
Thus, the corresponding code rate is
$R_k=\frac{\log{|\mathcal{W}_k|}}{n}$, where $|\mathcal{W}_k|$ denotes the
cardinality of $\mathcal{W}_k$.
The rate tuple
$(R_1,R_2,...,R_K)$ is said to be achievable if a sequence of
codebooks exists, such that the probability that each destination
decodes its message in error can be arbitrarily
small, by choosing long enough codewords.
The capacity region of
the network is the closure of the set of all achievable rates.
In Gaussian interference networks where the noise is additive white
Gaussian, the capacity region depends on the transmission powers of
sources, the noise powers and channel gains. Since the exact
capacity region is difficult to find, as a starting point one
can use the DoF region to
characterize/approximate the capacity/achievable rate region in
the high-SNR region (where interference is the dominant phenomenon
that degrades system performance).
The DoF region is defined as follows
\begin{equation}
\mathcal{D}=\left\{(d_1,...,d_K)\!\in\! \mathbb{R^+}|\exists
(R_1,...,R_K)\!\in\! \mathcal{C}(P), d_k=\!\lim_{P\rightarrow
\infty}\!\frac{R_k}{\log{P}},1 \leq \!k\! \leq K\right\},
\end{equation}
where $\mathcal{C}(P)$ denotes the capacity region, and $P$ is the
transmission power of each source.
The DoF can be seen as the pre-log
factor of the achievable rate and the DoF region describes how the capacity region expands as transmission power increases.

\section{Practical Challenges of Interference Alignment}
\label{Sec:Practical Challenges of Interference Alignment}
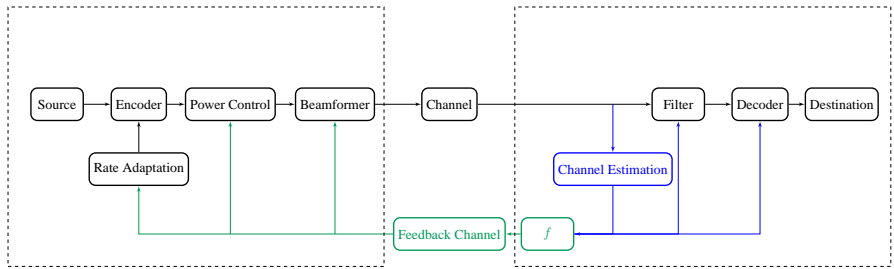
\begin{figure}
\centering
\scalebox{0.43}{
\begin{tikzpicture}[>=latex']
\tikzset{Box/.style={rectangle, draw, very thick, minimum width=1.6cm, minimum height=1cm, rounded corners=2mm}}
\Large

\node[Box] (S) at (-8,0) {Source};
\node[Box] (Enc) at (-5.5,0) {Encoder};
\node[Box] (PC) at (-2.7,0) {Power Control};
\node[Box] (BF) at (0.5,0) {Beamformer};
\node[Box] (CH) at (4,0) {Channel};
\node[Box] (Filter) at (11,0) {Filter};
\node[Box] (Dec) at (13.5,0) {Decoder};
\node[Box] (D) at (16,0) {Destination};
\draw [->,thick] (S.east) -- (Enc.west);
\draw [->,thick] (Enc.east) -- (PC.west);
\draw [->,thick] (PC.east) -- (BF.west);
\draw [->,thick] (BF.east) -- (CH.west);
\draw [->,thick] (CH.east) -- (Filter.west);
\draw [->,thick] (Filter.east) -- (Dec.west);
\draw [->,thick] (Dec.east) -- (D.west);

\node[Box,blue] (Ch_est) at (9,-2) {Channel Estimation};
\draw [->,thick,blue] (9,0) -- (Ch_est.north);

\node[Box,ForestGreen] (Ch_fb) at (7,-4) {$f$};
\draw [->,thick,blue] (Ch_est.south) |- (Ch_fb.east);
\draw [->,thick,blue] (Ch_fb.east) -| (Filter.south);
\draw [->,thick,blue] (Ch_fb.east) -| (Dec.south);

\node[Box,ForestGreen] (Fb_ch) at (4,-4) {Feedback Channel};
\draw [->,thick,ForestGreen] (Ch_fb.west) -- (Fb_ch.east);
\draw [->,thick,ForestGreen] (0.5,-4) -- (BF.south);
\draw [->,thick,ForestGreen] (-2.7,-4) -- (PC.south);
\node[Box] (RA) at (-5.5,-2) {Rate Adaptation};
\draw [->,thick,ForestGreen] (Fb_ch.west) -| (RA.south);
\draw [->,thick] (RA.north) -- (Enc.south);

\draw [dashed] (-9.5,-5) rectangle (2,3);

\draw [dashed] (6,-5) rectangle (17.5,3);
\end{tikzpicture} 
}
\caption{\footnotesize Transmitter and Receiver Structure.}
\label{Fig: system model}
\end{figure}

The structure of a canonical transmitter and receiver for the implementation of interference alignment is shown in Fig.~\ref{Fig: system model}.
At the transmitter side, there is an \textit{encoder} which encodes the messages to the corresponding codewords suitable for transmission over the channel.
The transmission can be enhanced by the adaptation of the transmitted signal according to the received channel state infomation feedback.
Specifically, in a class of communication systems that transmission powers are fixed and a maximum throughput is desired, the encoder may adapt transmission rate according to the estimate of the mutual information of the channel (computed by the \textit{rate adaptation} module).
On the other hand, in another class of
systems which desire fixed-rate transmission, \textit{power control} module should adjust transmitted power according to the channel state feedback to maintain mutual information of the channel larger than a certain level.
Each transmitter has a \textit{beamformer} which compute the proper signal for transmission over the channel according to the interference alignment concept.

At the receiver side, \textit{channel estimation} module computes the estimation of incoming channel gains.
These channel estimations can be used for recovering the transmitted message and computing the channel state information feedback signal.
The \textit{filter} module exploits estimated channel gains to recover the desired signal from interference signals according to the interference alignment concept.
The \textit{decoder} module decodes the message using an estimate of the incoming channel gain.
The feedback encoder module denoted by `$f$' in Fig. \ref{Fig: system model} computes the feedback signal according to the estimated channel gains. Also, there is a \textit{synchronizer} module at the receiver to synchronize the receiver and the transmitter. In the following, we will explain these parts in more detail.

\subsection{Channel Training for Interference Alignment}
\label{sec: channel training}
In practice, destinations can acquire CSI through a pilot-based channel training scheme.
For example, consider block fading channel model in which channel gains are constant within one fading block and change to independent values in the subsequent blocks. The length of each block coincides with the coherence time of channel denoted as $T$.

Terminals first need to learn the channels within each fading block, and next use the estimated channels to perform their transmission.
A pilot-assisted interference alignment scheme is proposed in \cite{Farhadi2014c} which perform these tasks. According to this scheme, transmission within each fading block is conducted in two phases: \textit{pilot transmission phase} and \textit{data transmission phase}.
These two phases have the duration of $T_{\tau} = \alpha T$ and $T_d = (1-\alpha)T$, respectively, where $\alpha\in[K/T,1]$  called \textit{channel sharing factor} is a design parameter.
A larger $\alpha$ leads to more accurate channel estimation at the expense of less time left for data transmission.
The transmission power of the pilot symbols in general can be different from that of the data symbols.
Let $P$, $P_d$, and $P_{\tau}$ denote the average transmission power, the average power of data symbols, and the average power of pilot symbols, respectively.
Source $\textnormal{S}_l$ $(l\in\{1,...,K\})$ sends $T_{\tau}$ known pilot symbols with power $P_{\tau}$.
Then, the destination $\textnormal{D}_k$ $(k\in\{1,...,K\})$ applies a minimum mean square error (MMSE) estimator to obtain an estimate of the channel gain.
As shown in Fig.~\ref{Fig: system model} this channel estimate is used at the receiver filter and the decoder to recover the desired message~\cite{Farhadi2013b}.

A more accurate estimation of the channel can be obtain by allocating more transmission power for training symbols which implies that a lower power is left for data transmission.
The achievable rate region by taking into account this noisy CSI is computed in \cite{Farhadi2014c}.
According to Proposition 2 in \cite{Farhadi2014c}, the optimum power allocation which maximizes the achievable rate region is
\begin{eqnarray}
P_{d,\textnormal{opt}}&=&\beta_\textnormal{opt} P,\nonumber\\
P_{\tau,\textnormal{opt}}&=&K\left({\left(1-\left(1-\alpha\right)\beta_\textnormal{opt}\right)}/{\alpha}\right)P,
\end{eqnarray}
where
\begin{eqnarray}
\label{eqn: optimum beta}
\beta_\textnormal{opt}=\frac{1}{1-\alpha}{\left(1+\sqrt{\frac{1+KP/(1-\alpha)}{1+PT/N_0}}\right)^{-1}}.
\end{eqnarray}

If $P\gg 1$, then the optimum power allocation factor is approximately
\begin{eqnarray}\label{optimum beta: large power}
\beta_\textnormal{opt}\approx
\frac{1/(1-\alpha)}{1+\sqrt{KN_0/T(1-\alpha)}}.
\end{eqnarray}

This result recommends that in large networks more power should be allocated to the channel training instead of the data transmission.
The intuition behind this result is that in large networks the performance loss due to imperfect interference alignment as a consequence of imperfect CSI becomes more important. Thus, it is recommended to allocate more power to pilot symbols instead of data symbols to acquire CSI more accurately.

%
%

\subsection{Channel State Information Feedback}

As we have discussed in the previous section, destinations can acquire CSI through a pilot-based channel training scheme.
The destinations then can send the estimated CSI to the sources via channel state feedbacks.
They can transmit either un-quantized CSI (analog feedback) or quantized CSI (digital feedback) via feedback channels.
In the following, we briefly review some key results for different type of feedback schemes.

\subsubsection{Analog Feedback}
The destinations can obtain an estimate of the incoming channels according to the scheme mentioned in Section \ref{sec: channel training}.
Then, they may transmit the analog value of the estimated channels over the feedback channel.
Let the function $f$ in Fig. \ref{Fig: system model} to be defined as $f(x)=x$, and assume error-free feedback channels.
According to Theorem 3 in \cite{Farhadi2014c}, in a single-input single-output (SISO) time-varying $K$-user network with coherence time $T$,
the achievable sum DoF~is
\begin{eqnarray}
d_{\sum}=\min\{K(1-K/T)/2,T/8\}.
\end{eqnarray}
This result is achieved when the number of the users selected to be active is
\begin{eqnarray}
K_{opt}=\min\{K,T/2\}.
\end{eqnarray}
This recommends that, in large networks $(K>T/2)$ when no CSI is a priori available at terminals,
first select a subset of the users and next perform channel training and interference alignment within the set of active users.

\subsubsection{Digital Feedback}
\label{sec: Digital Feedback}

Digital channel state feedback strategies in which each destination quantize the incoming channels and sends the index of the quantized channel over feedback channels has been studied in several works (e.g. \cite{Bolcskei2009,Krishnamachari2010,Kim2012}).
It is shown that the same DoF as when perfect CSI is available can be obtained, provided that the destinations a priori know channels and the feedback signals' rate is proportional to $\log{P}$, where $P$ is the transmission power of each source.
That is,
\begin{eqnarray}
N_f\sim \log{P},
\end{eqnarray}
where $N_f$ is the number of feedback bits.
It should be noticed that in practice where destinations do not know incoming channels a priori, part of the radio resources need to be allocated for channel training and a smaller DoF will be achievable (cf. \cite{Farhadi2014c}). This part is discussed in more detail in section \ref{perz1}.

\subsection{Iterative Interference Alignment}
\label{sec: Distributed Interference Alignment}
In this part we present an iterative algorithm referred to as \textit{leakage minimization algorithm} and an extension for it called \textit{Max-SINR} which both are proposed in \cite{Gomadam2011}.
Consider a three-user multiple-input multiple-output (MIMO) interference network where each terminal is equipped with $M$ antennas.
For presentation simplicity here we assume M to be even. The result when $M$ is odd is provided in \cite{Cadambe2008}.
It has been shown in \cite{Cadambe2008} that the achievable DoF of each source--destination pair is $\frac{M}{2}$.
To achieve this DoF, the transmitter-side beamformers and the receiver-side filters should be designed.
In the following, we will briefly present the algorithm proposed in \cite{Gomadam2011} to compute the beamformers and filters.
We assume each terminal can acquire only local channel side information, i.e. knowledge about the channels which are directly connected to it through out training of the channel.
Destination $\mathrm{D}_k$ can obtain $\mathbf{H}_{kl}$ and source $\mathrm{S}_k$ can obtain $\mathbf{H}_{lk}$ , $l\in\{1,2,3\}$, where $\mathbf{H}_{lk}$ is the forward channel from
$\mathrm{S}_k$ to $\mathrm{D}_l$ and ${\mathbf{H}}_{kl}^r$ is the reverse channel from
$\mathrm{D}_l$ to $\mathrm{S}_k$.
The iterative computation of the beamformers and the filters occur in the training phase.
After the convergence of the computed filters and beamformers to the interference alignment solutions, the data transmission starts.
Let $\mathbf{V}_k$ denote an $M\times \frac{M}{2}$ transmitter-side beamforming matrix where its
columns are the orthogonal basis of the transmitted signal space of $\mathrm{S}_k$.
The transmitted signal of
$\mathrm{S}_k$ is
\begin{eqnarray}
\mathbf{x}_k=\mathbf{V}_k\overline{\mathbf{x}}_k,
\end{eqnarray}
where each element of the $\frac{M}{2}\times 1$ vector
$\overline{\mathbf{x}}_k$ represents an independently encoded Gaussian codeword with power $\frac{2P_k}{M}$.
Each codeword is beamformed with the corresponding column of $\mathbf{V}_k$.

Let $\mathbf{U}_k$ be an $M \times \frac{M}{2}$ receiver-side filtering matrix whose
columns are the orthogonal basis of the desired signal subspace at
$\mathrm{D}_k$.
The filter output of $\mathrm{D}_k$ is
\begin{eqnarray}
\overline{\mathbf{y}}_k=\mathbf{U}_k^*\mathbf{y}_k,
\end{eqnarray}
where $\mathbf{y}_k$ is the received vector.
The transmitter-side beamforming matrices and the receiver-side filtering matrices should satisfy the following interference alignment conditions:
\begin{eqnarray}
\label{eqn: interefrence alignment conditions}
\mathbf{U}_k^*\mathbf{H}_{kj}\mathbf{V}_j&=&0,~\forall j\neq k:~j,k\in\{1,2,3\},\nonumber\\
\text{rank}(\mathbf{U}_k^*\mathbf{H}_{kk}\mathbf{V}_k)&=&\frac{M}{2},~\forall k: k\in\{1,2,3\}.
\end{eqnarray}
If global CSI is available, the beamforming and filtering matrices can be designed such that
these conditions are satisfied.
However, with the lack of global CSI
if we choose the beamformers and the filters randomly, with high probability
only the second condition in (\ref{eqn: interefrence alignment conditions}) will be satisfied.
Consequently, some interference remains at the destinations. The total power of interference at $\mathrm{D}_k$~is
\begin{eqnarray}
IF_k=\text{Tr}[\mathbf{U}_k^*\mathbf{Q}_k\mathbf{U}_k],
\label{IFK}
\end{eqnarray}
where
\begin{eqnarray}
\label{eq:interference covariance matrix}
\mathbf{Q}_k=\sum_{j=1,j\neq k}^3\frac{2P_j}{M}\mathbf{H}_{kj}\mathbf{V}_j\mathbf{V}_j^*\mathbf{H}_{kj}^*
\end{eqnarray}
is the covariance matrix of interference at $\mathrm{D}_k$.
Clearly, $IF_k=0$ only if the beamformers and the filters satisfy conditions in (\ref{eqn: interefrence alignment conditions}).
However, $\mathrm{D}_k$ can utilize the local channel side information to minimize this received interference power by optimizing its receiver-side filter.
Therefore, assuming that the beamformers are fixed, the receiver-side filter $\mathbf{U}_k$
is the solution of the following problem:
\begin{eqnarray}
\min_{\mathbf{U}_k,~\mathbf{U}_k\mathbf{U}_k^*=\mathbf{I}_{\frac{M}{2}}}IF_k.
\end{eqnarray}
The solution is given in \cite{Gomadam2011}:
\begin{eqnarray}
\label{eqn: filter DIA}
\mathbf{U}_k^d=\mathbf{\nu}^d[\mathbf{Q}_k],~d=1,...,\frac{M}{2},
\end{eqnarray}
where $\mathbf{U}_k^d$ denotes the $d$th column of $\mathbf{U}_k$ and $\mathbf{\nu}^d[\mathbf{A}]$ is the  eigenvector corresponding to the $d$th lowest eigenvalue of $\mathbf{A}$.

To optimize the transmitter-side beamformers the reciprocity of the channels can be exploited to obtain CSI at sources.
For instance, destinations can transmit training sequences over the \textit{reverse channels} (channels from destinations to sources) which are separated from the \textit{forward channels} (channels from sources to destinations) in time via time-division duplex (TDD).
The \emph{reciprocity} of the forward and reverse channels is assumed, i.e.
${\mathbf{H}}_{kl}^r=\mathbf{H}_{lk}^*$ ($\forall
l,k\in\{1,2,...,K\}$).
For this purpose, the destination $\mathrm{D}_k$ performs beamforming and $\mathrm{S}_k$ perform filtering in the reverse direction
with matrices ${\mathbf{V}}_k^r$ and
${\mathbf{U}}_k^r$, respectively.
These matrices can be selected to perform interference alignment in the reverse direction.
Since, the reciprocity holds, if we choose ${\mathbf{U}}_k^r=\mathbf{V}_k$ and
${\mathbf{V}}_k^r=\mathbf{U}_k$, the solutions of the
interference alignment problem in the reverse direction
are
equivalent to those in the forward direction.
In a similar way as in the forward direction, the reverse direction filters are computed as,
\begin{eqnarray}
({\mathbf{U}}_k^r)^d=\mathbf{\nu}^d[{\mathbf{Q}}_k^r],~d=1,...,\frac{M}{2},
\end{eqnarray}
where $\mathbf{Q}^r_k$ is the covariance matrix of interference at $S_k$ and is computed in a similar way as in \eqref{eq:interference covariance matrix} with the reverse direction channels and beamformers.
We can set the beamforming matrices in the forward direction as $\mathbf{V}_k={\mathbf{U}}_k^r$ and repeat this optimization procedure
until the beamforming matrices and the receiving filters converge.
The convergence of this algorithm is shown in \cite{Gomadam2011}.
Next, the sources beamform their data using the computed  beamforming matrices and the destinations decode their desired signals by applying associated filters.

An extension of this iterative algorithm is proposed in \cite{Gomadam2011} which instead of minimizing leakage at each destination tries to maximize signal-to-noise-plus-interference ratio (SINR) corresponding to each stream. This algorithm is referred to as \textit{Max-SINR algorithm} in the literature.
According to \textit{Max-SINR algorithm}, the receiver filtering vector $\mathbf{U}_k^d$ instead of the one in (\ref{eqn: filter DIA}) is selected to be a MMSE filter as follows
\begin{eqnarray}
\mathbf{U}_k^d=\frac{(\mathbf{Q}_k^d)^{-1}\mathbf{H}_{kk}\mathbf{V}_k^d}{\|(\mathbf{Q}_k^d)^{-1}\mathbf{H}_{kk}\mathbf{V}_k^d\|_2},
\end{eqnarray}
where
\begin{equation}\label{DIA def: Q}
\mathbf{Q}_k^d=\sum_{\substack{j=1}}^K\!\sum_{l=1}^{M/2}\!\frac{2P_j}{M}\mathbf{H}_{kj}\mathbf{V}_j^l\left(\mathbf{V}_j^{l}\right)^*\mathbf{H}_{kj}^*-\frac{2P_j}{M}\mathbf{H}_{kk}\mathbf{V}_k^d\left(\mathbf{V}_k^{d}\right)^*\mathbf{H}_{kk}^*+N_0\mathbf{I}_{M/2}.
\end{equation}
where $I_{M/2}$ is $\frac{M}{2} \times \frac{M}{2}$ identity matrix and $N_0$ is the noise power. Similarly,
Similarly, the transmitter beamforming vectors are updated in the reverse transmissions.
In the following sections, we will present test-bed implementation of algorithms which use Max-SINR algorithm for computing the transmitter and receiver filters.

\section{Review on Test-bed Implementation of Interference Alignment}
\label{Sec: Review on Test-bed Implementation of Interference Alignment}
The idea of squeezing aligned interference signals into half of the signal space and accessing the other half of the signal space for desired transmission in an interference network
is so tempting that a large body of works has been done since the introduction of interference alignment to implement this elegant approach, for instance look at ~\cite{GPK09,EPH10,BRA:12,ZM12,GRS11,MSS12,MAH13,MFZS:14,ZET:14a,ZET:14b}.

The first implementation of interference alignment is reported in~\cite{GPK09}.
A hybrid version of interference alignment combined with the successive interference cancellation (IAC) in a single carrier narrow-band MIMO wireless local area network (WLAN) is tested in this paper.
Several interference alignment and interference alignment-like approaches are tested in MIMO-OFDM interference channels in~\cite{EPH10}.
This paper specifically studies the effects of poor channel conditions on the performance of interference alignment.
Real-time implementation of different interference alignment scenarios are performed on different test-beds like the ones in~\cite{GRS11,MSS12,MAH13}.
\cite{GRS11} identifies practical issues that degrades the interference alignment performance such as channel estimation errors or collinearity between the desired signal and interference subspaces while in \cite{MAH13}, in Vienna MIMO test-bed (VMTB), the typical delay is measured.

Implementation of interference alignment in frequency domain over measured channels is considered in~\cite{BRA:12} where the different variants of interference alignment are compared with frequency planning scenarios. The significant superiority of interference alignment performance in terms of average sum rate is reported at high SNRs in this paper.

In all previously mentioned papers, the effects of hardware impairments are ignored.
In \cite{ZM12}, the effects of such impairments on the performance of interference alignment and coordinated multi-point
\footnote{CoMP is an approach similar to interference alignment with this main difference that in CoMP all the sources know the information to be transmitted to
all destinations.} (CoMP)
is investigated on \emph{KTH four-multi} test-bed and a measurement based model for signal-to-interference-noise-and-distortion ratio (SINDR) is suggested.
Implementation of interference alignment with power control and interference alignment with compressed feedback are also studied on the same test-bed in~\cite{MFZS:14} and~\cite{ZET:14a}, respectively. These two approaches are further explained in the rest of this chapter.
Table below summarizes some of the works that have been done in the test-bed implementation of interference alignment.
\begin{landscape}
\begin{table}
\label{table:review}
	\begin{tabular}{|p{1cm}|p{1.2cm}|p{1.5cm}|p{2cm}|p{1.75cm}|p{0.8cm}|p{1cm}|p{1.6cm}|p{0.7cm}|
p{2.5cm}|p{2cm}|}
	\hline
	Reference&alignment dimension&implemented algorithms & number of user pairs & channels &
	DL/UL& antennas per node & hardware & $f_c$ (GHz)
& synchronization &Measurement environment \\
	\hline \hline
	\cite{GPK09} & Space &\begin{tabular}[l]{@{}l@{}}IA in DL\\ IAC in UL \end{tabular} & 2 and 3 & flat-fading MIMO &
	DL and UL & 2 &\begin{tabular}[l]{@{}l@{}}USRP (one per\\  node) \end{tabular} &--
& --
& indoor office \\
	\hline
	\cite{EPH10} & Space & variants of IA  & 3 & MIMO-OFDM &
	DL &2 & NI PXI-1045 &2.4
& \begin{tabular}[l]{@{}l@{}}high precision OCXO\\ software correction \\ trigger signal  \end{tabular}&indoor and outdoor\\
	\hline
	\cite{ZM12}& Space & MAX-SINR & 3 & MIMO-OFDM &
	DL & 2 & USRP &2.49
&\begin{tabular}[l]{@{}l@{}}10 MHz CLK \\ NMEA \\ PPS \end{tabular} &indoor office \\
	\hline
	\cite{BRA:12} & Frequency & variants of IA & 3 & measured SISO-OFDM &
	DL&1&LTE BS &2.7
& N/A &urban macrocell \\
	\hline
	\cite{GRS11}& Space & IA & 3 & MIMO-OFDM &
	DL & 2 & Lyrtech  & 5
& \begin{tabular}[l]{@{}l@{}}40 MHz OSC\\ hardware trigger  \end{tabular}  &indoor office  \\
	\hline
	\cite{MSS12}& Space & IA & 3 & MIMO-OFDM &
	DL & 2 & USRP  & 2.4
&\begin{tabular}[l]{@{}l@{}}10 MHz CLK\\ PPS  \end{tabular}  &indoor office  \\
	\hline
	\cite{MAH13}& Space & IA & 3 (1 physical and 2 virtual Rx) & MIMO-OFDM &
	DL & 4 & VMTB  &2.503
&\begin{tabular}[l]{@{}l@{}}rubidium CLK\\ GPS  \end{tabular}  &\begin{tabular}[l]{@{}l@{}}outdoor to indoor\\ indoor to outdoor  \end{tabular}  \\
	\hline
\cite{MFZS:14}& Space & IA with PC & 3 & MIMO-OFDM &
	DL & 2 & USRP  & 2.4
&\begin{tabular}[l]{@{}l@{}}10 MHz CLK\\ NMEA \\PPS  \end{tabular}  &indoor office  \\
	\hline
\cite{MSS12}& Space & IA with compressed feedback & 3 & MIMO-OFDM &
	DL & 2 & USRP  & 2.4
&\begin{tabular}[l]{@{}l@{}}10 MHz CLK \\ NMEA \\ PPS \end{tabular}  &indoor office  \\
	\hline
	\end{tabular}
\label{table:review}
\end{table}

\end{landscape}

\section{KTH Four-multi Test-bed Setup}
\label{Sec: KTH four-multi Test-bed Setup}

\emph{KTH four-multi} is a USRP-based wireless communication test-bed consisting of several stationary and movable multi-antenna nodes.
A software framework accompanies the hardware setup of the test-bed which facilitates the rapid testing of multi-antenna schemes (see \url{http://fourmulti.sourceforge.net/}).
In the following, the hardware and software structure of the test-bed is described.

\subsection{Hardware Setup}
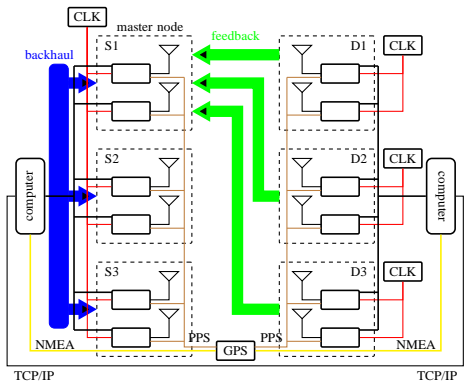
\begin{figure}
\begin{center}
\scalebox{0.5}{

\begin{tikzpicture}
\large
\tikzset{USRP/.style={rectangle, draw, very thick, minimum width=2*\unit, minimum height=\unit,inner sep=0pt,rounded corners=1pt}}
\tikzset{node/.style={rectangle, draw, thick, dashed, minimum width=5*\unit, minimum height=5*\unit,inner sep=0pt,rounded corners=1pt}}
\tikzstyle{feedback}=[line width=1mm,draw=green,-triangle 60,postaction={draw, line width=3mm, shorten >=4mm, -}]
\tikzstyle{backhaul}=[line width=1mm,draw=blue,-triangle 60,postaction={draw, line width=3mm, shorten >=4mm, -}]
\tikzset{backhaul2/.style={rectangle, draw=blue, fill=blue, minimum width=1*\unit, minimum height=14*\unit,inner sep=0pt,rounded corners=4pt}}
\tikzset{comp/.style={rectangle, draw, very thick, minimum width=4*\unit, minimum height=1.5*\unit,inner sep=0pt,rounded corners=4pt}}

\definecolor{darkgray}{rgb}{0.3,0.3,0.3}
\definecolor{lightgray}{rgb}{0.8,0.8,0.8}
\def \unit {5mm}
\def \dist {11*\unit}
\def \w {1.6}

\def\SourceAntenna{%
   -- + (\unit,0) -- +(\unit,\unit) -- + (\unit*1.5,\unit*1.5) -- +(\unit*0.5,\unit*1.5)
-- + (\unit,\unit)
}
\def\DestinationAntenna{%
   -- + (-\unit,0) -- +(-\unit,\unit) -- + (-\unit*1.5,\unit*1.5) -- +(-\unit*0.5,\unit*1.5)
-- + (-\unit,\unit)
}

\foreach \y in {0,1,2}{
\node[USRP] (A\y1) at (0,\y*3) {};
\draw[thick] ($(A\y1)+(\unit,0)$) \SourceAntenna;
\node[USRP] (A\y2) at ($(A\y1)+(0,2*\unit)$) {};
\draw[thick] ($(A\y2)+(\unit,0)$) \SourceAntenna;
\node[node] (S\y) at ($(A\y1)+(\unit*0.7,\unit*1.5)$) {};
}
\node[USRP,draw=none] at ($(A01)+(-\unit,\unit*3.5)$) {S3};
\node[USRP,draw=none] at ($(A11)+(-\unit,\unit*3.5)$) {S2};
\node[USRP,draw=none] at ($(A21)+(-\unit,\unit*3.5)$) {S1};
\node[USRP,draw=none] at ($(A21)+(\unit,\unit*4.5)$) {master node};

\foreach \y in {0,1,2}{
\node[USRP] (D\y1) at (\dist,\y*3) {};
\draw[thick] ($(D\y1)-(\unit,0)$) \DestinationAntenna;
\node[USRP] (D\y2) at ($(D\y1)+(0,2*\unit)$) {};
\draw[thick] ($(D\y2)-(\unit,0)$) \DestinationAntenna;
\node[node] (D\y) at ($(D\y1)+(-\unit*0.7,\unit*1.5)$) {};
}
\node[USRP,draw=none] at ($(D01)+(\unit,\unit*3.5)$) {D3};
\node[USRP,draw=none] at ($(D11)+(\unit,\unit*3.5)$) {D2};
\node[USRP,draw=none] at ($(D21)+(\unit,\unit*3.5)$) {D1};

\draw [feedback] ($(D2)+(-\unit*2.5,1.5*\unit)$)--($(S2)+(\unit*2.5,1.5*\unit)$);
\draw [feedback] ($(D1)+(-\unit*2.5,0)$)--++(-\dist*0.1,0) --++ (0,\unit*6) -- ($(S2)+(\unit*2.5,0)$);
\draw [feedback] ($(D0)+(-\unit*2.5,0)$)--++(-\dist*0.2,0) --++ (0,\unit*10.5) -- ($(S2)+(\unit*2.5,-1.5*\unit)$);
\node[USRP,draw=none,green] at ($(D2)+(-4.8*\unit,2.5*\unit)$) {feedback};

\draw [backhaul] ($(S2)+(-\unit*4,0)$)--($(S2)+(-\unit*2.5,0)$);
\draw [backhaul] ($(S1)+(-\unit*4,0)$)--($(S1)+(-\unit*2.5,0)$);
\draw [backhaul] ($(S0)+(-\unit*4,0)$)--($(S0)+(-\unit*2.5,0)$);
\node[backhaul2] (B) at ($(S1)+(-\unit*4.5,0)$) {};
\node[USRP,draw=none,blue] at ($(S1)+(-5*\unit,7.5*\unit)$) {backhaul};

\foreach \y in {0,1,2}{
\node[USRP] (CLK\y) at ($(D\y)+(4*\unit,2*\unit)$){CLK};
\draw [thick,red] (CLK\y)--++(0,-1.5*\unit)--(D\y2);
\draw [thick,red] (CLK\y)--++(0,-3.5*\unit)--(D\y1);
}
\node[USRP] (CLK3) at ($(S2)+(-3*\unit,3.5*\unit)$){CLK};
\foreach \y in {0,1,2}{
\foreach \x in {1,2}{
\draw [thick,red] (CLK3)--++(0,-3*\unit-2*\unit*6+\y*\unit*6-2*\unit*2+\x*\unit*2)--(A\y\x);
}
}

\foreach \y in {0,1,2}{
\foreach \x in {1,2}{
\draw [very thick] ($(A\y\x)+(-1*\unit,0.4*\unit)$)--++(-2*\unit,0);
\draw [very thick] ($(D\y\x)+(\unit,0.4*\unit)$)--++(\unit,0);
}
}
\draw [very thick] ($(A22)+(-3*\unit,0.4*\unit)$)--++(0,-14*\unit);
\draw [very thick] ($(D22)+(2*\unit,0.4*\unit)$)--++(0,-14*\unit);
\node[comp,rotate=90] (comp1) at ($(S1)-(6*\unit,0)$){computer};
\node[comp,rotate=-90] (comp2) at ($(D1)+(6*\unit,0)$){computer};
\draw [very thick] (comp1)--++(2.3*\unit,0);
\draw [very thick] (comp2)--++(-3.3*\unit,0);
\draw [thick] (comp1)--++(-1.22*\unit,0)--++(0,-9*\unit)--++(2*\dist+2*\unit,0)--++(0,9*\unit)--(comp2);
\node[USRP,draw=none] at ($(comp1)+(0.2*\unit,-9.5*\unit)$) {TCP/IP};
\node[USRP,draw=none] at ($(comp2)+(-0.2*\unit,-9.5*\unit)$) {TCP/IP};

\node[USRP] (GPS) at (\dist/2,-0.7*\unit){GPS};
\draw[thick,brown] ($(GPS)+(\unit,0.2*\unit)$)--++(1.7*\unit,0)--++(0,\unit*14.3);
\foreach \y in {0,1,2}{
\foreach \x in {1,2}{
\draw[thick,brown] ($(D\y\x)-(\unit,0.2*\unit)$)--++(-1.8*\unit,0);
}
}
\node[USRP,draw=none] at ($(D01)+(-3.6*\unit,0)$) {PPS};
\draw[thick,brown] ($(GPS)+(-\unit,0.2*\unit)$)--++(-1.7*\unit,0)--++(0,\unit*14.3);
\foreach \y in {0,1,2}{
\foreach \x in {1,2}{
\draw[thick,brown] ($(A\y\x)+(\unit,-0.2*\unit)$)--++(1.8*\unit,0);
}
}
\node[USRP,draw=none] at ($(A01)+(3.6*\unit,0)$) {PPS};

\draw[very thick,yellow] (GPS)--++(\dist-0.2*\unit,0)--(comp2);
\draw[very thick,yellow] (GPS)--++(-\dist+0.2*\unit,0)--(comp1);

\node[USRP,draw=none] at ($(D01)+(4*\unit,-0.4*\unit)$) {NMEA};
\node[USRP,draw=none] at ($(A01)+(-4*\unit,-0.4*\unit)$) {NMEA};

\end{tikzpicture}

}
\end{center}
\caption{Network structure of KTH four-multi test-bed.}
\label{fig:network_structure}
\end{figure}

The current version of the test-bed consists of six nodes where three of them are fixed and take the role of transmitting sources
while the other three are movable receiving destinations.
All the nodes are equipped with two vertically polarized dipole antennas spaced 20 cm apart which is 1.6 times of the carrier's wavelength.
Twelve Ettus Research USRP N210 (see \url{www.ettus.com}) are used to govern the twelve antennas in the network.
The source USRPs are equipped with the standard Ettus XCVR2450 RF dautherboards while
the destination USRPs use custom boards to achieve sufficient noise figure and dynamic range.
The output signal of each source USRP is amplified by a ZRL-2400LN power amplifier.
Two Linux computers control all the USRPs in the network.
The network structure of four-multi test-bed is illustrated in Fig.~\ref{fig:network_structure}.

The network is designed to work at 2.49 GHz center frequency with 12 MHz bandwidth.
Synchronization of the network is performed in three levels, namely time, frequency and transmit-receive synchronizations.
The time and transmit-receive synchronizations are done by means of a pulse-per-second (PPS) signal (0-5 V, 1 Hz square wave) and
a national marine electronics association (NMEA) signal (an ASCII protocol that provides hour-minute-second time), respectively.
Both signals are generated by an EM406A GPS module and distributed through the network.
The frequency synchronization is also performed by helps of 10 MHz reference clocks (CLK).
All the source's local oscillators are locked to the same clock while a separate clock is provided for each of the destinations.
In a real implementation the same synchronization would
be achieved using common control and synchronization channels
(cellular systems) or from the burst preambles (wireless local area networks).
In a system with interference alignment, transmitter will in any case
need some kind of back-haul to provide a common time reference
and disperse scheduling decisions.

\subsection{Software Setup}
The four-multi software framework has been developed in C++ (see \url{http://fourmulti.sourceforge.net/}).
It run on two Linux computers separately. One of the computers controls the three source nodes while the other one controls the three destination nodes connected to them via Ethernet connections.
The sources\rq{} computer generates the transmitted frames and feeds them to the source nodes while the destinations\rq{} computer process the received frames at the destination nodes.
A TCP/IP connection between the source and the destination computers provides the feedback links. Backhaul communication among the source nodes  is also implemented by the help of TCP/IP connections between the source computer and the source nodes.

The framework contains a toolbox for coding and modulation ({\tt AMC} and {\tt OFDM1})
which was used in the implementations of the next two sections.
The modulation and coding toolbox includes an LDPC channel encoder/decoder, a QAM modulator/demodulator and an OFDM modulator/demodulator.
The specifications of these built-in functions is summarized in Table~\ref{table: toolbox specifications}.


\begin{table}[t!]
\centering
\caption{KTH four-multi modulation and coding toolbox specifications.}
\label{table: toolbox specifications}
\begin{tabular}[t!]{|l|c||l|c|}
\hline
\multicolumn{2}{|c||}{\tt OFDM1}&\multicolumn{2}{|c|}{\tt AMC}\\
\hline
FFT length & 80 & Coding rates & $\frac{1}{2},\frac{5}{8},\frac{3}{4}$\\
\hline
Cyclic prefix length & 12 & Codeword length & 1520 \\
\hline
Number of null subcarriers & 42 & QAM modulation orders & 4, 16, 64, 256 \\
\hline
Subcarrier spacing (KHz) & 312.5 & & \\
\hline
\end{tabular}
\end{table}


\section{Test-bed Implementation of the Iterative Transceiver Filter Design and Power Control}
\label{Sec: Test-bed Implementation of the Iterative Transceiver Design and Power Control}

%

We, in this section,  first  present an iterative algorithm for joint transceiver filter design and power control proposed in \cite{Farhadi2013} and then explain how this algorithm is implemented on KTH four-multi test-bed and finally present measurement results.

\subsection{Iterative Transceiver Filter Design and Power Control Algorithm}
In an interference network, each user can affect the received SINR at its corresponding destination through the
choice of beamforming and receiving filters as well as the transmitting power. Considering single stream transmissions,
the received SINR at the $k$th destination is computed as
\begin{equation}
\label{eq:SINR}
{\rm SINR}_k \big({ \mathbf{u}_k,\mathbf{v}_k,P_k}\big) = \frac{P_k \mathbf{u}_k^* \mathbf{H}_{kk} \mathbf{v}_k
\mathbf{v}_k^* \mathbf{H}_{kk}^* \mathbf{u}_k }{ IF_k + N_0}
\end{equation}
where $IF_k$ is given by (\ref{IFK}).
Given the beamforming and receiving filters, from Equation \eqref{eq:SINR} the minimum transmitting power required to maintain a fixed rate $R_{tar,k}$
(assuming that the interference can be regarded as Gaussian noise)
at $k$th destination can be found as
\begin{align}
\label{eq:standard_interference_function}
P_k = \beta_k \big({ \mathbf{u}_k,\mathbf{v}_k, \gamma_{k}}\big)
:= \frac{ \gamma_k \left(IF_k + N_0\right)}{\mathbf{u}_k^* \mathbf{H}_{kk} \mathbf{v}_k
\mathbf{v}_k^* \mathbf{H}_{kk}^* \mathbf{u}_k },
\end{align}
where $\gamma_k = 2^{R_{tar,k}} -1$ is the minimum required SNR in an AWGN channel.
The total power of the interference, $IF_k$ which appears  in the nominator is a function of transmitting powers at all the interfering transmitters.
It means that increasing the power of one source causes higher level of interference at non-corresponding destinations and
therefore the other sources need to transmit with higher power as well.
In~\cite{Yates:95}, it is shown that if the relation between the powers $P_k$, $k = 1,\dots,K$
satisfies a set of conditions then either there is
an unique solution for the powers that can be found iteratively starting from any initial value or no solution exists.
It is easy to prove that Equation~\eqref{eq:standard_interference_function} meets the conditions in~\cite{Yates:95}
and therefore an optimal transmitting powers can be found distributively after a sufficient number of iterations.

Inspired by Max-SINR algorithm and the aforementioned power control algorithm,
an iterative transceiver design and power control algorithm was proposed in \cite{Farhadi2013}.
A brief version of the algorithm is presented on the next page for the sake of completeness.
The algorithm is composed of three update phases in each iteration such that the
receiving filters, transmission powers and beamforming filters are sequentially updated.
The receiving and beamforming filters are optimized to deliver the maximum SINR at the destinations
in the forward communication direction and at the sources in the reverse direction, respectively according to the
concept of Max-SINR algorithm.
On the other hand, in the power update phase, the powers are set to the minimum values needed for maintaining a fixed rate communication. The transmission power is upper bounded by $P_{\max}$.
This algorithm assumes that accurate CSI is obtained at terminals. An extensions of the algorithm when terminals have access to only noisy CSI is proposed in \cite{Farhadi2013a}.

\begin{algorithm}[t]
\caption{Transceiver Filter Design and Power Control \cite{Farhadi2013}}
\label{alg: iterative IA and PC algorithm}
\begin{algorithmic}
\STATE \textbf{Initialize:} $\mathbf{v}_{k}(0)$, $P_{k}(0)$,  $k\in\{1,...,K\}$, $n=1$.
\REPEAT
\STATE \textbf{Update receiver filtering vector:}
\begin{eqnarray}
\mathbf{u}_k(n) = \arg \max_{\mathbf{u}_k} \left\{ {\rm SINR}_k \left(\mathbf{u}_k,\mathbf{v}_k(n-1), P_k(n-1) \right) \right\}
\end{eqnarray}
\STATE \textbf{Update transmission power:}
\begin{eqnarray}
\scalebox{1.1}
{$
P_{k}(n)=\min \left\{\beta_k \left( { \mathbf{u}_k(n), \mathbf{v}_k(n-1), {\gamma}_{k}}\right) ,P_{\max}\right\}
$}
\end{eqnarray}
\STATE \textbf{Reverse the communication direction:}
\hspace{1cm}
$
\left\{
        \begin{array}{l}
            \mathbf{u}_k^{r}(n-1) = \mathbf{v}_k(n)\\
            \mathbf{H}_{ij}^{r} = \mathbf{H}_{ji}^* \hspace{0.5cm} i,j\in \{1,\dots,K\} \hspace{1.7cm}  \\
            P_k^{r}(n-1) = P_F
        \end{array}
\right.
$

\STATE \textbf{Update transmitter beamforming vector:}
\begin{eqnarray}
\mathbf{u}_k^r(n) = \arg \max_{\mathbf{u}_k^r} \left\{ {\rm SINR}_k^r \left(\mathbf{u}_k^r,\mathbf{v}_k^r(n-1), P_k^r(n-1) \right) \right\}
\end{eqnarray}
\STATE Set beamforming vector  $\mathbf{v}_{k}(n)={\mathbf{u}}_k^r(n)$,
and $n=n+1$.
\UNTIL{$n=N$}
\end{algorithmic}
\end{algorithm}


\subsection{Transmitted Frame Structure}

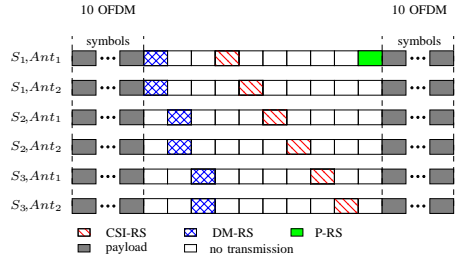
\begin{figure}
\begin{center}
\scalebox{0.7}{


\begin{tikzpicture}
\definecolor{darkgray}{rgb}{0,0,1}
\definecolor{lightgray}{rgb}{1,0,0}
\def \unit {2.8mm}
\def \w {1.6}
\tikzset{symbol/.style={rectangle, draw, minimum width=\w*\unit, minimum height=\unit,inner sep=0pt}}
\tikzset{legend/.style={rectangle, draw, minimum width=\w*\unit/2, minimum height=\unit/2,inner sep=0pt}}
\node(empty) at (\w*\unit,0){};

\foreach \y in {0,2*\unit,4*\unit,6*\unit,8*\unit,10*\unit}
{
\node[symbol,fill=gray] (d1) at (0,\y) {};
\draw [fill] ($(d1)+(\w*\unit-0.25*\w*\unit,0)$) circle  (0.02);
\draw [fill] ($(d1)+(\w*\unit,0)$) circle  (0.02);
\draw [fill] ($(d1)+(\w*\unit+0.25*\w*\unit,0)$) circle  (0.02);
\node[symbol,fill=gray] (d2) at ($(d1)+2*(empty)$) {};
\node[symbol] (dem_p1) at  ($(d2)+(empty)$) {};
\node[symbol] (dem_p2) at  ($(dem_p1)+(empty)$) {};
\node[symbol] (dem_p3) at  ($(dem_p2)+(empty)$) {};
\node[symbol] (csi_p1) at  ($(dem_p3)+(empty)$) {};
\node[symbol] (csi_p2) at  ($(csi_p1)+(empty)$) {};
\node[symbol] (csi_p3) at  ($(csi_p2)+(empty)$) {};
\node[symbol] (csi_p4) at  ($(csi_p3)+(empty)$) {};
\node[symbol] (csi_p5) at  ($(csi_p4)+(empty)$) {};
\node[symbol] (csi_p6) at  ($(csi_p5)+(empty)$) {};
\node[symbol] (csi_p7) at  ($(csi_p6)+(empty)$) {};
\node[symbol,fill=gray] (d3) at  ($(csi_p7)+(empty)$) {};
\draw [fill] ($(d3)+(\w*\unit-0.25*\w*\unit,0)$) circle  (0.02);
\draw [fill] ($(d3)+(\w*\unit,0)$) circle  (0.02);
\draw [fill] ($(d3)+(\w*\unit+0.25*\w*\unit,0)$) circle  (0.02);
\node[symbol,fill=gray] (d4) at  ($(d3)+2*(empty)$) {};
\node[symbol] (CSI) at  ($(csi_p6)-(\y*\w*0.5,0)$) {};
\draw [pattern color=red, pattern=north west lines] ($(CSI)-(\w*\unit/2,\unit/2)$) rectangle ($(CSI)+(\w*\unit/2,\unit/2)$);
}
\node[symbol] (DM1) at  ($(dem_p1)+(0,0)$) {};
\draw [pattern color=blue, pattern=crosshatch] ($(DM1)-(\w*\unit/2,\unit/2)$) rectangle ($(DM1)+(\w*\unit/2,\unit/2)$);
\node[symbol] (DM2) at  ($(dem_p1)+(0,-2*\unit)$) {};
\draw [pattern color=blue, pattern=crosshatch] ($(DM2)-(\w*\unit/2,\unit/2)$) rectangle ($(DM2)+(\w*\unit/2,\unit/2)$);
\node[symbol] (DM3) at  ($(dem_p1)+(\w*\unit,-4*\unit)$) {};
\draw [pattern color=blue, pattern=crosshatch] ($(DM3)-(\w*\unit/2,\unit/2)$) rectangle ($(DM3)+(\w*\unit/2,\unit/2)$);
\node[symbol] (DM4) at  ($(dem_p1)+(\w*\unit,-6*\unit)$) {};
\draw [pattern color=blue, pattern=crosshatch] ($(DM4)-(\w*\unit/2,\unit/2)$) rectangle ($(DM4)+(\w*\unit/2,\unit/2)$);
\node[symbol] (DM5) at  ($(dem_p1)+(2*\w*\unit,-8*\unit)$) {};
\draw [pattern color=blue, pattern=crosshatch] ($(DM5)-(\w*\unit/2,\unit/2)$) rectangle ($(DM5)+(\w*\unit/2,\unit/2)$);
\node[symbol] (DM6) at  ($(dem_p1)+(2*\w*\unit,-10*\unit)$) {};
\draw [pattern color=blue, pattern=crosshatch] ($(DM6)-(\w*\unit/2,\unit/2)$) rectangle ($(DM6)+(\w*\unit/2,\unit/2)$);

\draw[fill=green]  ($(csi_p7)+(-\w*\unit/2,-\unit/2)$) rectangle ($(csi_p7)+(\w*\unit/2,\unit/2)$);

\node  (txt6) at (-2*\w*\unit,0) {\scriptsize{$S_3\!,\!Ant_2$}};
\node  (txt5) at (-2*\w*\unit,2*\unit) {\scriptsize{$S_3\!,\!Ant_1$}};
\node  (txt4) at (-2*\w*\unit,4*\unit) {\scriptsize{$S_2\!,\!Ant_2$}};
\node  (txt3) at (-2*\w*\unit,6*\unit) {\scriptsize{$S_2\!,\!Ant_1$}};
\node  (txt2) at (-2*\w*\unit,8*\unit) {\scriptsize{$S_1\!,\!Ant_2$}};
\node  (txt1) at (-2*\w*\unit,10*\unit) {\scriptsize{$S_1\!,\!Ant_1$}};

\draw [dashed] ($(0,0)-(0.5*\w*\unit,0)$) --++ (0,11.5*\unit);
\draw [dashed] ($(3*\w*\unit,0)-(0.5*\w*\unit,0)$) --++ (0,11.5*\unit);
\draw [dashed] ($(13*\w*\unit,0)-(0.5*\w*\unit,0)$) --++ (0,11.5*\unit);
\draw [,dashed] ($(16*\w*\unit,0)-(0.5*\w*\unit,0)$) --++ (0,11.5*\unit);
\node[rectangle,minimum width=2.8*\w*\unit, align=center]  at ($(txt1)+(1.35,0.6)$) { \scriptsize{10 OFDM}\\ \scriptsize{symbols}};
\node[rectangle,minimum width=2.8*\w*\unit, align=center]  at ($(txt1)+(7.2,0.6)$) { \scriptsize{10 OFDM}\\ \scriptsize{symbols}};

\node[legend] (leg1)  at  ($(0,0)-(0,0.5)$) {};
\draw [pattern color=red, pattern=north west lines] ($(leg1)-(\w*\unit/4,\unit/4)$) rectangle ($(leg1)+(\w*\unit/4,\unit/4)$);
\node[legend,fill=gray] (leg2)  at  ($(0,0)-(0,0.8)$) {};
\node[legend] (leg3) at  ($(0,0)-(-2,0.5)$) {};
\draw [pattern color=blue, pattern=crosshatch]  ($(leg3)-(\w*\unit/4,\unit/4)$) rectangle ($(leg3)+(\w*\unit/4,\unit/4)$);
\node[legend] (leg4) at  ($(0,0)-(-2,0.8)$) {};
\node[legend,fill=green] (leg5) at  ($(0,0)-(-4,0.5)$) {};

\node[rectangle,minimum width=2.8*\w*\unit, align=left] (leg-txt2) at ($(leg2)+(0.8,0)$) { \scriptsize payload};
\node[rectangle,minimum width=2.8*\w*\unit, align=left] (leg-txt1) at ($(leg1)+(0.8,0)$) { \scriptsize CSI-RS};
\node[rectangle,minimum width=2.8*\w*\unit, align=left] (leg-txt3) at ($(leg3)+(0.8,0)$) { \scriptsize DM-RS};
\node[rectangle,minimum width=2.8*\w*\unit, align=left] (leg-txt4) at ($(leg4)+(1.15,0)$) { \scriptsize{no transmission}};
\node[rectangle,minimum width=2.8*\w*\unit, align=left] (leg-txt5) at ($(leg5)+(0.6,0)$) { \scriptsize{P-RS}};

\end{tikzpicture}

}
\end{center}
\caption{Transmitted frame structure in IA-PC scheme.}
\label{fig:frame_structure}
\end{figure}

The air interface of the network is designed based on OFDM modulation using Four-multi's modulation and coding toolbox.
Coding rate of $1/2$ and 16QAM modulation was chosen for transmitting fixed-rate data streams though the air interface.
The transmitted frame structure is depicted in Fig.~\ref{fig:frame_structure}. In our experiment, each frame consists of 20 payload symbols and either two or three reference signals (RS) (i.e. pilot symbols).  The payload symbols are \emph{concurrently} transmitted from all the sources while each stream is beamformed with its corresponding beamforming filter instructed by Algorithm~\ref{alg: iterative IA and PC algorithm}. Although the overhead of pilot symbols is significant in this case, we note that the number of payload symbols could be larger depending on the coherence time of the indoor channel. In the implementation in Section VII there are in fact thousands of payload symbols with no additional pilots. Thus the overhead could be reduced to one percent for the environment in Fig.~\ref{fig:measurement environment map}.

Three types of RS are employed in the network, which are referred to as \textit{channel state information RS (CSI-RS)}, \textit{demodulation RS (DM-RS)} and \textit{power RS (P-RS)}. During the pilot transmission all the sub-carriers of the OFDM symbol is filled with known QAM symbols. We next explain each type of these reference signals:
\begin{itemize}
\item CSI-RS:  The received noisy CSI-RS at the destinations are exploited to estimate the corresponding channel matrices to enable execution of Algorithm~\ref{alg: iterative IA and PC algorithm}. The CSI-RS are transmitted \emph{orthogonally}; i.e., one CSI-RS is transmitted from each transmit antenna in the network while the other antennas are silent. To enhance the quality of the channel estimations the CSI-RS are scaled such that the associated QAM symbol has the maximum transmit power $P_{\max}$.
\item  DM-RS: The DM-RS are used to compute the effective channel by taking into account the transmit and receive filters. Therefore they need to be stream-dedicated and be processed by the same pre-coder as the payload symbols of the corresponding stream. In this way, their power is not fixed and is set by the power control algorithm.
\item   P-RS:  Algorithm~\ref{alg: iterative IA and PC algorithm} is constructed to select the minimum possible transmission power to minimize the interference at the destinations. This hence reduces the power of DM-RS and may lead to a poor estimation of cross-channels, which is not favorable. To tackle this problem, P-RS is introduced where the amplitudes of the DM-RS are scaled after the pre-coder by a scaling factor $\alpha$. In each frame, the scaling factor is computed as
\begin{equation}
\alpha = \sqrt{\frac{P_{\max}}{\max_{k,j}P_{k,j}}},
\end{equation}
where $P_{k,j}$  is the transmit power of the $j$th sub-carrier of $k$th source. The destinations therefore need to get informed about the scaling factor. This is achieved by having node 1 (the master node) repeat its first CSI-RS but now scaled with $\alpha$. This enables all destinations to make robust estimate of $\alpha$ (it is assumed that all destinations can hear node 1). The factor $\alpha$ is also quantized into a discrete set of values to avoid that $\alpha$ introduces estimation errors.
\end{itemize}

\subsection{Measurement results}
\begin{figure}
\begin{center}
\includegraphics[width=0.35\columnwidth,trim=2cm 3cm 2cm 3cm]{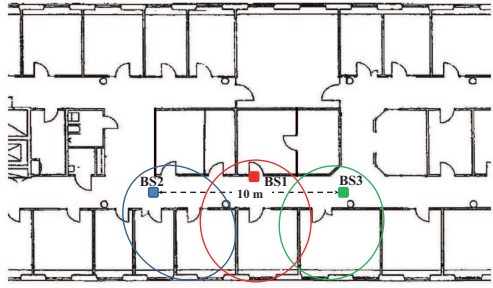}
\end{center}
\caption{Measurement environment map.}
\label{fig:measurement environment map}
\end{figure}

The test-bed measurement was performed in KTH signal processing department which floor map is illustrated in Fig.~\ref{fig:measurement environment map}. The measurement environment is categorized as an indoor office.
In this experiment only non-line-of-sight (non-LOS) scenarios
were investigated by placing source  and destination nodes  in the corridor and inside the nearby offices, receptively.
The receive antenna gains also decreased by connecting 10 dB attenuators to them in order to avoid saturation of receive power amplifiers.

The measurement was done in 100 batches.
In each batch a random placement of the destination nodes in the area marked by colored circles in the figure were measured.
The signals transmitted according to two different schemes were measured sequentially in each batch.
In the first scheme, referred to as \textit{noPC}, the iterative interference alignment was implemented
according to~\cite{Gomadam2011} for benchmarking.
In the second scheme, referred to as \textit{PC}, transmission powers and beamforming filters were computed according to Algorithm~\ref{alg: iterative IA and PC algorithm} and MMSE receiving filters were applied at the destinations. Each scheme was run with 28 frames inter-spaced 0.15 seconds.
The statistics of the first frames of both schemes were not taken into account since there is no feedback information at these frames.

High power may push the terminals' power amplifier to work in their non-linear region.
Non-linearities in the transmit-receive chain degrades the performance of the system by introducing \emph{distortion noise} into the system.
Distortion noise is usually modeled as a Gaussian noise whose power increases by increasing the transmission power of the source nodes~\cite{Schenk08}.
In order to make sure that the reduction of transmit power is not the only cause for the performance improvement,
four different levels of transmission power were tested in noPC scheme that is each 7 frames were transmitted with a different power.
\begin{table}[t!]
\centering
\caption{Measurement results.}
	\begin{tabular}{ | c |c | c | c | c| c|}
		\hline
		scheme &\multicolumn{4}{|c|}{noPC}& PC\\
		\hline \hline
		Average power~(dBm) & -12.9 & -3.4 & 1.1 & 7.1 & -1 \\
		\hline
		FER & 0.6856 &   0.1700 &   0.0528 &   0.0561 & 0.0071 \\
		\hline
		BER & 0.0815  &  0.0124  &  0.0020   & 0.0030 &  $2.2 \times 10^{-4}$ \\
		\hline
		Average SINR (dB) & 10.9 & 20 & 24.3& 26.7 &18.5\\
		\hline
	\end{tabular}
\centering
\label{table:noPCvsPC}
\end{table}

Table~\ref{table:noPCvsPC} shows the average performance of the two schemes for the 100 measurement batches.
In this measurement the PC scheme's target SINR $\gamma_k$ was set to $18$ dB for all the users.
The performances were compared in the sense of bit-error rate (BER) and transmitted power.
The table shows that the PC scheme has the lowest BER, although its average transmit power is
much lower than the noPC scheme with the best BER performance.

\begin{figure}
\begin{center}
\scalebox{0.35}{\input{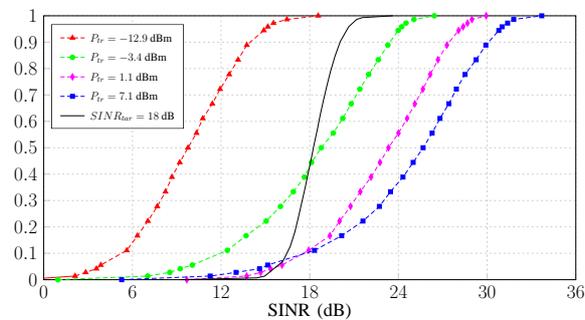}}
\end{center}
\caption{\footnotesize Empirical CDF of SINR. The solid line represents the PC scheme and the dashed lines denote the noPC scheme with different transmit powers ($P_{tr}$).}
\label{fig:CDF of SINR}
\end{figure}

Empirical cumulative distribution function (CDF) of the received SINR for two schemes is plotted in Fig~\ref{fig:CDF of SINR}. This plot reveals the reason for the low BER of PC scheme despite its low transmit power.
The received SINRs in this scheme are concentrated around the target value while in the noPC scheme they are distributed over a wider range.
Having SINR higher than the target value while the transmit rate is fixed leads to the waste of energy
and on the other hand SINRs lower than the target increase the probability of error and therefore the BER.


\begin{figure}
\begin{center}
\scalebox{0.35}{
\input{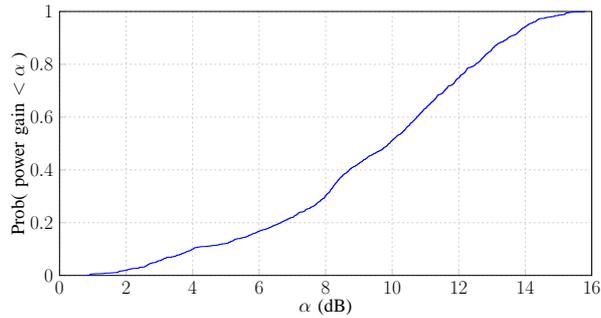}
}
\end{center}
\caption{\footnotesize Emperical CDF of transmit power saving gain.}
\label{fig:CDF of gain}
\end{figure}

\emph{Power saving gain} at each frame is computed as the ratio of the transmit power in noPC scheme with 7.1 dBm average power and the total power transmitted in the PC scheme with 18 dB target SINR.
Fig.~\ref{fig:CDF of gain} shows the empirical CDF of the power saving gain. As the empirical CDF implies, implementation of Algorithm~\ref{alg: iterative IA and PC algorithm} in PC scheme leads to at least 4 dB gain in $90\%$ of the measurements. Fig.~\ref{fig:CDF of gain} also shows that in $10\%$ of the measurements gains higher than 13 dB was observed.

The benefit of power control in the PC scheme is in fact two-fold. By decreasing the transmit power,
while retaining the target SINR, not only less interference is received at the destinations but also the distortion noise
due to transceiver impairments decreases.

\section{Test-bed Implementation of the Interference Alignment with Compressed Feedback}
\label{Sec: Test-bed Implementation of the Interference Alignment with Compressed Feedback}
\newcommand{\diag}{\mathop{\mathrm{\mathbf{diag}}}}

In this section we describe the implementation of interference using limited
digital feedback (see Section \ref{sec: Digital Feedback}).
Rather than using a uniform quantizer, we use a modified
form of the MIMO matrix compression of the IEEE802.11ac standard.
The propagation scenario and testbed is the same as the one used
in Section \ref{Sec: Test-bed Implementation of the Iterative Transceiver Design and Power Control}
i.e. a system with three ($K=3$), $2\times 2$ MIMO links.
However, the transmission power is higher than in previous sections and the system is therefore
limited by interference and hardware impairments rather than thermal noise \cite{ZET:14a,ZET:14b}.
A second difference from the previously presented results in Section
\ref{Sec: Test-bed Implementation of the Iterative Transceiver Design and Power Control},
is the performance criterion.
Rather than using BER we here use transmission rate as the criterion.
The transmission rate is obtained by probing each link with ten different
coding and modulation schemes (MCS). The rate is determined by finding the MCS
with the highest rate which does not incur any frame errors.

The section is organised as follows. Section \ref{perz1} describes the feedback
compression
of the CSI assuming a system with $K$ links and $M$
antennas in each source and destination node,
while the measurement results are
presented in Section \ref{perz2}.

\subsection{Compression of IEEE802.11ac and adaptation to interference
alignment}
\label{perz1}

The feedback scheme described in the standard IEEE802.11ac resembles the feedback method for
slowly time-varying single-user MIMO channels presented in \cite{ROH:07}. In this scheme a singular value
decomposition (SVD) of the MIMO channel ${\mathbf{H}}$ (for a certain sub-carrier)
is first performed as

\begin{equation}
{\mathbf{H}} = {\mathbf{W}}{\mathbf{\Lambda}}{\mathbf{F}^*}.
\end{equation}
The destination then feeds back, in compressed form,
the complex unitary matrix $\mathbf{F}$ 
and the real diagonal matrix $\mathbf{\Lambda}$, while the complex unitary matrix
${\mathbf{W}}$ does not need to be fed back.
To realize this we may imagine that each destination pre-multiplies its received signal with
${\mathbf{W}}^*$. The effective channel then becomes $\mathbf{\breve{H}}={\mathbf{\Lambda}}{\mathbf{F}^*}$
which is a channel which can be reconstructed by the sources. In practice, the destination
will not pre-multiply its received signal with ${\mathbf{W}}^*$, since almost every receive
technique is invariant to such a linear unitary transform.

In the case of centralised interference alignment, knowledge of the channels
between all sources and destinations
is required to obtain all the beamformers $\mathbf{V}_k$.
We assume here that sources are connected to a common back-bone
for exchange of CSI and synchronization.
If we directly apply the compression scheme presented above
for the single-user case so that each destination $D_k$,
compresses the matrices ${\mathbf{H}}_{k,l}$ 
for $l \in \{ 1,\ldots,K \}$ it is clear that there is
no single unitary matrix to be applied at the destination to transform all
involved K channels into
$\mathbf{\breve{H}}_{k,l}=\mathbf{\Lambda}_{k,l} \mathbf{F}^*_{k,l}$
such that all channels can be obtained from
the compressed feedback from user $k$.

To overcome this problem, in the
system implementation we have based the feedback from destination $k$ on the big matrix,
$\mathbf{H}^{[k]}$ obtained by concatenating the channel matrices ${\mathbf{H}}_{k,l}$
for $l=1,\ldots,K$ column-wise. Thus in this case $\mathbf{H}^{[k]}$, 
is computed as,

\begin{equation}
\mathbf{H}^{[k]} = [ \mathbf{H}_{k,1} , \ldots, \mathbf{H}_{k,K} ].
\end{equation}

Thus $D_k$ now feeds back a compressed version of the right-hand side
eigenvectors of $\mathbf{H}^{[k]}$ i.e. the real diagonal matrix $\mathbf{\Lambda}^{[k]}$
and the complex unitary matrix $\mathbf{F}^{[k]}$ of the SVD

\begin{equation}
\mathbf{H}^{[k]} = {\mathbf{W}}^{[k]}  {\mathbf{\Lambda}^{[k]}} (\mathbf{F}^{[k]})^{*}.
\end{equation}

The size of matrix  $\mathbf{\Lambda}^{[k]}$ is $M \times M$ while $\mathbf{F}^{[k]}$
is $KM \times M$. Just as in the single-link case we can now imagine
that the destination pre-multiplies its received signal with
$(\mathbf{W}^{[k]})^*$.  The transmitter side is then able
to obtain knowledge of the effective channel
$\mathbf{\breve{H}}^{[k]}={\mathbf{\Lambda}^{[k]}}({\mathbf{F}^{[k]}})^*$.
From this combined channel the constituent sub-channels can be
pulled out from the corresponding columns e.g. the second
channel ${\breve{\mathbf{H}}_{k,2}}$ corresponds to columns
$M+1,\ldots,2M$ of $\mathbf{\breve{H}}^{[k]}$.
The reconstructed channels can then be used in place of the actual
channels in any transmit beamformer and receive filter algorithm.

The IEEE 802.11ac feedback compression scheme starts by rotating the phase
of the columns of ${\mathbf{F}^{[k]}}$
in order the last row of this matrix to have real and positive elements.
These phase rotations do not
need to be sent to the sources, since these rotations only amount to a phase
rotation of
the signals which can be
undone at the destination.
In the next step, ${\mathbf{F}^{[k]}}$,
is multiplied by a diagonal matrix to have the first column
become real and positive as

\begin{equation}
{\mathbf{F}^{[k]}} \longleftarrow {\mathbf{F}^{[k]}} \diag(\exp(j\phi_{1,1}),
\ldots,\exp(j\phi_{KM-1,1}),1).
\end{equation}

These angles $\phi_{1,1},\ldots,\phi_{KM-1,1}$ are then quantised uniformly, each with $b_{\phi}$ bits
. Then, real-valued
Givens rotations are utilised to successively zero out the second to the
$KM$-th element of the first
column of ${\mathbf{F}^{[k]}}$ . For the latter rotations the
angles $\psi_{2,1},\ldots,\psi_{KM,1}$ are used. The angles lie between 0
and $\pi/2$ and are quantised uniformly with $b_{\psi}$ bits.
This procedure is repeated in a similar fashion
for the remaining columns of
 ${\mathbf{F}^{[k]}}$.
More details are presented in \cite{ROH:07} and
\cite{POR:13} as well as in the Matlab$/$Octave functions available at \url{http://people.kth.se/~perz/packV/}.

Since we use OFDM there is one
${\bf H}^{[k]}$ matrix for each user and sub-carrier.
In the IEEE 802.11ac standard, the parameter $N_{\textnormal{g}}$
is defined. This parameter determines the
frequency domain granularity of the feedback.
If $N_{\textnormal{g}}=1$, the feedback of
 ${\mathbf{F}^{[k]}}$ is done on every sub-carrier.
If $N_{\textnormal{g}}=2$, the feedback is only done on every other sub-carrier.
In the standard, the values
1 , 2 and 4 have been defined for  $N_{\textnormal{g}}$ .
In our measurements the values 8 , 16 and 38 have also been considered,
since this would significantly reduce
the number of feedback bits. The angle resolution
parameters $b_{\phi}$ and $b_{\psi}$ are defined in
the IEEE 802.11ac standard as either
$b_{\phi}=5$ and $b_{\psi}=7$ or
$b_{\phi}=7$ and $b_{\psi}=9$. In the presented results only the latter
value pair is used.
The total number of bits required to feed back the ${\mathbf{F}^{[k]}}$ matrix
is given by

\begin{equation}
n_{\textnormal{b}} = ( (2 K M - 1) M - M^2)( b_{\phi}+ b_{\psi} )/2
\end{equation}

The number of bits can be reduced by a further $(K-1)b_{\phi}$ bits.
We can see this by dividing the ${\mathbf{F}^{[k]}}$ into sub-matrices
of size $M \times M$ as

\begin{equation}
({\mathbf{F}^{[k]}})^* = [ ({\mathbf{F}_1^{[k]}})^*,\ldots,({\mathbf{F}_{K}^{[k]}})^*].
\end{equation}

Since the the signals from source, $l$, only propagates through sub-matrix ${\mathbf{F}^{[k]}}_l$,
we may freely multiply sub-matrices $1,\ldots,K-1$ by one phasor each. By doing so, we
may set $\phi_{1,1},\ldots,\phi_{1+(K-2)M,1}$ to zero, and thus we do not need to transmit them over the feedback
channels. The last sub-matrix can not be rotated since then the last row of  ${\mathbf{F}^{[k]}}$
would no longer be real and positive.

The elements of ${\mathbf{\Lambda}^{[k]}}$  are divided by the noise standard
deviation. The so obtained value can be regarded as the SNR of a
corresponding stream.
The reporting of these SNRs are done separately
per stream, and is done in two steps. In the
first step, the average SNR in dB over the whole frequency band, i.e.
for all singular values of the
narrowband channels for which SNR is reported, is fed back to $S_k$.
This value is uniformly
quantised with 8 bits in the range from 10 dB to 53.75 dB.
In the second step, the difference in
dB between the SNR of the reported sub-carrier and the average SNR is
computed and is uniformly quantised with 4 bits in the range from 8 dB to 7 dB. In $S_k$ , the
SNRs of all the reported sub-carriers are first reconstructed.
Then, linear interpolation (in dB) is
deployed to obtain the SNR for all the sub-carriers for which  ${\mathbf{F}^{[k]}}$ is fed back. With these two entities
at hand, the channel matrices $\mathbf{H}^{[k]}$
are reconstructed and used to compute the desired
pre-coders. For the sub-carriers where there is no feedback available, the pre-coding of the nearest
reported sub-carrier is utilized, i.e. no matrix interpolation method is used.

A practical problem which occurred during the early experimentation
was that the SNR
sometimes exceeded 53.75dB. This happened due to the high transmission power
and short range. First this was handled by reporting the maximum value
53.75dB whenever this happened.
However, this resulted in making the reconstructed channel at the transmitter high-rank although the estimated channel
at the receiver were in fact low-rank
- which was very detrimental to the performance.
To circumvent this problem, an offset is subtracted from SNR values when this condition
happens.

\subsection{Results}
\label{perz2}

\begin{figure}
\centerline{
\scalebox{0.6}{
\input{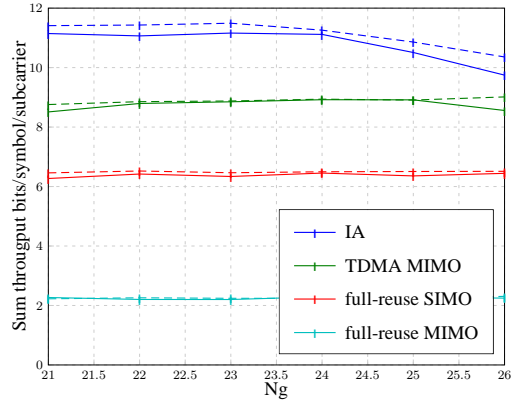}
}
}
\caption{\footnotesize Measured sum throughput. The dotted
line are the results without compression}
\label{perzfig1}
\end{figure}

In addition to 
interference alignment
\footnote{By interference alignment we are here refering to the
modified form of interference alignment known as Max-SINR, see Section \ref{sec: Distributed Interference Alignment}.
However, measurements we have performed has shown that the
performance difference between the original interference alignment and Max-SINR
is neglible in our scenario.},
we also run
the reference scheme single-user MIMO.
In this scheme
two spatial streams are transmitted from each source
using its two transmitter antennas.
Two versions of the single-user MIMO is used.
In one version one source is active at a time.
This version is naturally called TDMA.
In another version of single-user MIMO all links are active at the same time.
This scheme will work well if the inter-link interference is weak. We call
this approach \textit{full-reuse MIMO}.
Finally, we also include a scheme called \textit{full-reuse SIMO}. In this approach
all sources transmit at the same time but using only a single-antenna at the sources.
The difference between this scheme and interference alignment is the beamforming at the sources
and the channel state feedback. All other aspects of the signal processing are identical.
The performance of the system were investigated in 43 different locations for each of the three destination terminals.
Half of the locations were in the corridor
and half in the adjacent rooms, see Figure~\ref{fig:measurement environment map}.
In each location the performance was investigated
with the frequency domain granularity parameter, $N_{\textnormal{g}}$ set
sequentially to the values $1,2,4,8,15,38$. No person or object was moving in the
environment and all schemes were run in rapid sequence thus
making it possible to reliably compare the results. The results are shown
in Figure \ref{perzfig1}. The dashed lines are the results obtained
without the feedback compression while the solid lines are with
compression (although the frequency domain granularity is still applied).
The results show negligible performance loss from the quantization.
However, using a too high frequency domain granularity,
$N_{\textnormal{g}}>8$ or 2.5MHz, do incur some loss.
Note that the result for full-reuse MIMO is very poor showing that the
interference is strong. The gain of interference alignment over the best reference scheme
(TDMA MIMO) is 27\%. We may also note that interference alignment provides a gain over
full-reuse SIMO which proves that the transmitter beamforming
is making a difference.

When the channel is changing the performance of interference alignment will inevitably
be degraded due the channel mismatch between the channel at the
feedback time and the channel at the actual transmission.
The amount of degradation depends on the time delay and the
level of movement in the environment. In \cite{ZET:14b}
using measurements we showed that the throughput of interference alignment dropped 6.4\% when
two passers-by were walking in the measurement environment and the
feedback delay interval was 23 ms. With this delay interval
the overhead of performing the feedback scheme above with
 $N_{\textnormal{g}}=8$ is 2.5\%, assuming that the feedback
scheme of IEEE802.11ac is used.

The channel may also change due to the existence of the users at the destinations.
This effect was studied by measuring the performance
with and without a user located close to the destination nodes.
The results were obtained using eight destination positions.
The performance of interference alignment dropped 4.9\% while the performance
of single-user MIMO was unaffected.

These results show that interference alignment can still deliver a net performance
gain of sum throughput some 15\% over single-user MIMO in a WiFi scenario
with  three access points and three users, even when feedback overheads
and mobility is taken into account.

\section{Conclusion}
\label{Sec: Conclusion}
We have reviewed the concept of interference alignment, with theoretical results
regarding power and time allocation between training and payload data transmission,
and previous works within experimentation on interference alignment.
We further present implementation efforts addressing the combination of iterative interference
alignment and power control and using compressed channel state feedback based on a modified
form of the MU-MIMO feedback scheme of IEEE802.11ac.
Our experimental results show that the iterative interference alignment and power control scheme is able to provide better FER/BER performance (16QAM code rate 0.5), than an implementation without power control.
The results using the modified IEEE802.11ac feedback scheme, show that interference aligment
can bring an improvement in throughput when considering the loss of bandwidth needed
for feedback of the channel state information even for channels with realistic indoor mobility.

\section{Acknowledgments}
\label{Sec: Acknowledgemnts}
The research leading to these results has received funding from
Swedish Foundation for Strategic Research (SSF) under RAMCOORAN project and
ACCESS Linnaeus Center under graduate course
wireless experimentations.
The measurements were performed within the framework of the HIATUS project.
The project HIATUS acknowledges the financial support of the Future and Emerging
Technologies (FET) programme within the Seventh Framework Programme for Research of
the European Commission under FET-Open grant number:~265578.

\bibliographystyle{IEEEtr}
\addcontentsline{toc}{chapter}{Bibliography}
\bibliography{IEEEfull,References_Thesis_Hamed,References_PerZ,References_NimaM}
\end{document}